# Electronic structure and charge-density wave transition in monolayer VS$_2$

Hyuk Jin Kim[1], Byoung Ki Choi[1], In Hak Lee[1,2], Min Jay Kim[1,3], Seung-Hyun Chun,[4] Chris Jozwiak,[5] Aaron Bostwick,[5] Eli Rotenberg,[5] and Young Jun Chang,[1,3,*]

## Author affiliations

[1]*Department of Physics, University of Seoul, Seoul 02504, Korea,*

[2]*Center for Spintronics, Korea Institute of Science and Technology, University of Seoul, Seoul 02792, Korea*

[3]*Department of Smart Cities, University of Seoul, Seoul 02504, Korea*

[4]*Department of Physics, Sejong University, Seoul 05006, Korea*

[5]*Advanced Light Source (ALS), E. O. Lawrence Berkeley National Laboratory, Berkeley, California 94720, USA*

\* Corresponding authors: yjchang@uos.ac.kr

# Abstract

Vanadium disulfide ($VS_2$) attracts elevated interests for its charge-density wave (CDW) phase transition, ferromagnetism, and catalytic reactivity, but the electronic structure of monolayer has not been well understood yet. Here we report synthesis of epitaxial 1T $VS_2$ monolayer on bilayer graphene grown by molecular-beam epitaxy (MBE). Angle-resolved photoemission spectroscopy (ARPES) measurements reveal that Fermi surface with six elliptical pockets centered at the M points shows gap opening at low temperature. Temperature-dependence of the gap size suggests existence of CDW phase transition above room temperature. Our observations provide important evidence to understand the strongly correlated electron physics and the related surface catalytic properties in two-dimensional transition-metal dichalcogenides (TMDCs).

## I. INTRODUCTION

Complex phases in two-dimensional (2D) layered transition metal dichalcogenides (TMDCs) attract great interest for fundamental phenomena, such as band-gap transition, charge-density wave (CDW), and superconductivity, as well as their potential energy applications, such as solar energy harvesting and efficient catalytic electrode alternatives [1–4]. Due to inherent 2D layered geometry, TMDCs have shown diverse CDW phases in metallic TMDCs, such as $Ta(S,Se)_2$, $Nb(S,Se)_2$, $V(S,Se,Te)_2$ [5–9]. Those metallic TMDCs also recently show potential performances as phase switching device for nonvolatile memory[10], CDW-based oscillator[11], and photodetector [12], and catalytic electrode for hydrogen evolution reaction[13].

Vanadium dichalcogenides, especially, attract much attention due to their intriguing properties, such as CDW, ferromagnetism, and surface catalytic behavior. Bulk $VS_2$ shows CDW phase transition at 304 K as detected by previous electron diffraction pattern, temperature-dependent resistivity and magnetic susceptibility, and nuclear magnetic resonance measurements [14,15]. However, angle-resolved photoemission spectroscopy (ARPES) study of bulk $VS_2$ concluded the absence of Fermi surface (FS) nesting during the CDW transition [7]. On the other hand, bulk $VSe_2$ has a CDW phase (105 K) with $4 \times 4 \times 3$ periodicity which is attributed to FS nesting [8], while bulk $VTe_2$ has strong CDW phase (482 K) with $4 \times 4 \times 3$ periodicity [9]. Recent extensive studies have revealed that $VSe_2$ and $VTe_2$ exhibits transition from 3D nesting vector to 2D one, depending on the thickness, while the detailed role of thickness, interface, or stoichiometry remained controversial [16–19] In fact, huge attention was focused on vanadium dichalcogenides due to its theoretically calculated ferromagnetism in their monolayer (ML) form [20,21]. While extensive theoretical results predicted both 1H and 1T phase for ML $VX_2$ (X=S, Se, Te) with spin-polarized band structures, most experiments show the 1T phase ML with non-magnetic band structures in the case of $VSe_2$ and $VTe_2$. Detection of room temperature ferromagnetic signal in ML $VSe_2$ was understood as a result of either non-stoichiometric defects or interface-driven effect [22–24]. Another interest is surface catalytic property found at the surface of ultrathin $VS_2$ and $VSe_2$, whose reactivity due to metallic conductivity and atomic distortion shows quite promising results [25–27]. Therefore, it is important to study fundamental electronic structure properties of the single-crystalline ultrathin films with combination of molecular-beam epitaxy (MBE) and ARPES.

However, the MBE growth of vanadium chalcogenides have been confined with Se and Te,

because high vapor pressure of sulfur is difficult to be incorporated with the ultra-high vacuum (UHV) instruments. As an alternative of elemental sulfur, metal sulfides, such FeS and $FeS_2$, have been recently attempted to obtain high quality epitaxial sulfide thin films for limited number of sulfide compounds, such as $MoS_2$, $WS_2$, $NbS_2$, and $TaS_2$ [28,29]. Still, there is only limited number of reports on the epitaxially grown $VS_2$ ML [30], in which temperature dependent electronic structure have not been understood yet.

Here, we successfully prepared ML $VS_2$ on graphene substrates by using MBE system. ML $VS_2$ shows similar in-plane lattice parameter to its bulk value and is epitaxially aligned to the graphene with small tensile strain. ML $VS_2$ film shows a FS map with six-fold symmetry, well aligned to the band structure of graphene. The electronic band structure is compared with the theoretical band structure of freestanding ML $VS_2$ with 1T phase. Temperature dependence of gap sizes are analyzed for understanding the CDW in this ML $VS_2$.

## II. EXPERIMENTS

ML $VS_2$ was grown on bilayer graphene (BLG) on 4H-SiC (0001) using a home-built UHV MBE with a base pressure of $1.0\times10^{-9}$ torr [31]. After outgassing at 650°C for a few hours, the substrates were annealed up to 1300°C for 6 min for preparation of BLG on SiC substrates while monitoring reflection high-energy electron diffraction (RHEED) images. High-purity V (99.8%) rod and FeS (99.98%) powder were used as vanadium and sulfur sources, and simultaneously evaporated by an electron-beam evaporator (EFM3, Omicron GmBH) and a Knudsen cell (Effucell Co.), respectively. According to previous studies, only sulfur atoms are evaporated at 800-900°C from the FeS source, which is also confirmed with increase of $S_2$ signal (Z = 64) by using a residual gas analyzer (RGA) installed in the MBE [28,32]. During the growth, the substrate temperature was maintained at 250°C, followed with post-annealing at 350°C for 30 min. The growth process was monitored with *in situ* RHEED, and the growth rate was 20 min per ML. After growth, the as-grown samples are examined for its topography by using atomic force microscopy (AFM, XE-100, PSIA) in ambient condition. For the following UHV characterizations, the sample was additionally capped with 100 nm of amorphous Se at room temperature. The sample is transferred through air to a UHV ARPES analysis system. Then, the Se capping layer was thoroughly evaporated by annealing the sample to 300°C for several hours in an UHV before the subsequent characterizations.

The ARPES and XPS measurements were carried out in the micro-ARPES end-station at the MAESTRO facility beamline 7.0.2 at the Advanced Light Source, Lawrence Berkeley National Laboratory. The ARPES system was equipped with a VG Scienta R4000 electron analyzer. The lateral size of the synchrotron beam was estimated to be between 30 and 50 μm. We used photon energies of 115 eV to map Brillouin zone (BZ) and fine temperature-dependent energy-dispersion curves. The Fermi energy ($E_F$) is calibrated with polycrystalline gold foil attached at the side of sample. The total energy resolution was better than 20 meV, and the angular resolution was 0.1°. For XPS measurements, the photon energy of 350 eV was used with energy resolution of 85 meV.

## III. RESULTS AND DISCUSSION

Fig. 1(a) display a schematic model of ML $VS_2$ stacked on a BLG/SiC substrate. The crystal structure consists of a flat hexagonal plane of vanadium atoms sandwiched between two sulfur layers, forming a 1T structure with octahedral symmetry, similar to the other ML vanadium dichalcogenides, $VSe_2$ and $VTe_2$ [33,34]. Figure 1(b,c) shows RHEED images of BLG and ML $VS_2$ with same measurement parameters, respectively. The BLG shows several main streaks with additional diffraction spots due to different scatterings from complex interfaces between

BLG and SiC [17]. When ML VS$_2$ layer is formed, the features of BLG disappear and only vertical streak lines of (1×1) RHEED pattern emerge, which is indicative of smooth film growth with uniform crystallinity on top of the graphene lattice. Lack of additional diffraction lines indicate absence of additional surface reconstruction such as (2×2) RHEED patterns in case of distorted 1T' structures [35]. To characterize the in-plane lattice parameters from such ultrathin film, we compare the horizontal spacing of the RHEED streaks between the graphene and VS$_2$ film, as shown in Fig. 1(d). From the lattice constant of graphene (2.46 Å), in-plane lattice parameter of the ML VS$_2$ is estimated to be 3.26 Å, which is similar to the bulk lattice (3.23 Å) as well as the thin film lattice (3.20 Å) [15,36]. The minimum lattice mismatch is estimated to be 1.5% (tensile strain) for the match between 3 VS$_2$ unit cells and 4 graphene unit cells. The BLG substrate is a suitable choice for stacking van der Waals heterostructures and for investigating their electronic structure due to the chemical inertness and electrical conductivity with small FS area [33,37]. We also examined topography of the as-grown ML VS$_2$ film by using *ex situ* AFM. Figure 1(e) shows smooth film surface with small ML-height islands dispersed on the graphene substrate with step structures are barely visible (white arrows). The root-mean-square value of surface roughness is as small as 0.84 nm, which is smooth enough for the following photoemission measurements. Note that scanning tunneling microscopy measurements are desirable for atomic scale investigation, while our ambient AFM topography shows smooth surface with limited lateral resolution [30].

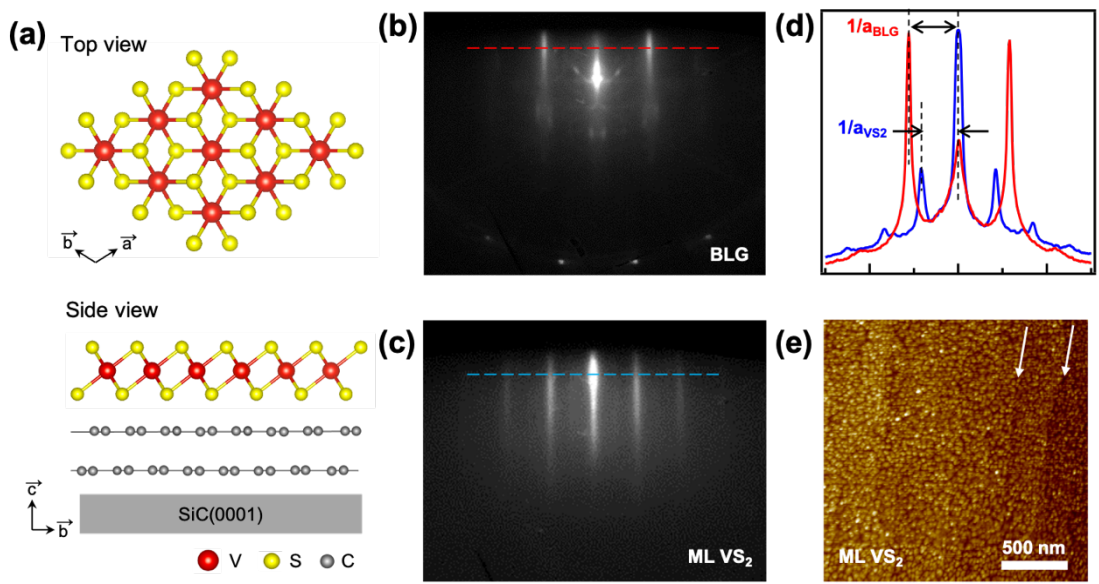

Figure 1. (a) Top- and side-view schematics of ML VS$_2$ on BLG. RHEED patterns of (b) BLG and (c) ML VS$_2$. (d) Horizontal line profiles of RHEED patterns on BLG (red) and ML VS$_2$ (blue). (e) Topographic AFM image of the ML VS$_2$.

We characterize chemical state of the ML VS$_2$ by using XPS spectra. Core-level spectra of the VS$_2$ film shows mostly V and S signals in addition to Si and C signals from the substrate, as shown in Fig. 2(a), after evaporating the amorphous Se layer. We note that the very small peak at ~54 eV may be originated from either Se 3$d$ from capping layer or Fe 3$p$ from FeS source. Since the vapor pressure of Fe is known to be very low at the given source temperature [28], so the small peak is assigned to the remanent Se even after the decapping procedure. The V 3$p$ peak in Fig. 2(b) with a shoulder signal in higher binding energy can be divided into three peaks centered at 37.48 eV (V1), 38.55 eV (V2) and 40.71 eV (V3). The peak V1 indicates that vanadium derived from sulfide, while the other shoulder peaks (V2 and V3) may be originated from non-stoichiometric phases with sulfur vacancies, selenium or oxygen defects [30,38]. Similarly, the narrow scan near the S 2$p$ peak in Fig. 2(c) displays multiple peaks, composed of four spin-orbit splitting doublets between S $2p_{1/2}$ and $2p_{3/2}$ with an energy difference of 1.18 eV and an amplitude ratio of 1:1.92. Two different S $2p_{3/2}$ (S $2p_{1/2}$) peaks of top (S1) and bottom (S2) sulfur sites in the single layer are located at 160.91 eV (162.11 eV) and 161.5 eV (162.7 eV) due to the different neighboring conditions, such as top surface and bottom interface to BLG, respectively, as previously reported in the ML $VS_2$ layer [30]. In addition, the other two S signals may be derived from C-S-C bonding (S3) in the S-doped graphene at 162.13 eV (163.31 eV) and non-stoichiometric contribution (S4) at 163.67 eV (164.85 eV), respectively [39–41]. We note that further systematic studies of core level states for various thicknesses is desired to monitoring evolution of each bonding states. We also analyzed the carbon core level spectrum, C 1$s$, consisted with 6 different peaks (C1-C6) in Fig. 2(d). The peak C1 is attributed to carbon from the SiC substrate at 283.8 eV. Stack of BLG and buffer layer give rise to three peaks (C2-C4). The $sp^2$ bonds in the BLG and buffer layer contributes the peak C2 at 284.25 eV and the peak C4 at 285.1 eV, respectively. The $sp^3$ bonds between the buffer layer and SiC gives the peak C3 at 284.65 eV.[42–44]. The remaining two small peaks may originate from other carbon compounds, such as C-S-C (C5, 285.65 eV) and C-O (C6, 286.2 eV) [41,45]. We note that majority of the film consists of well-ordered phase of $VS_2$ on BLG and is ready for electronic structure analysis, although graphene is slightly doped according to the peaks S3 and C5.

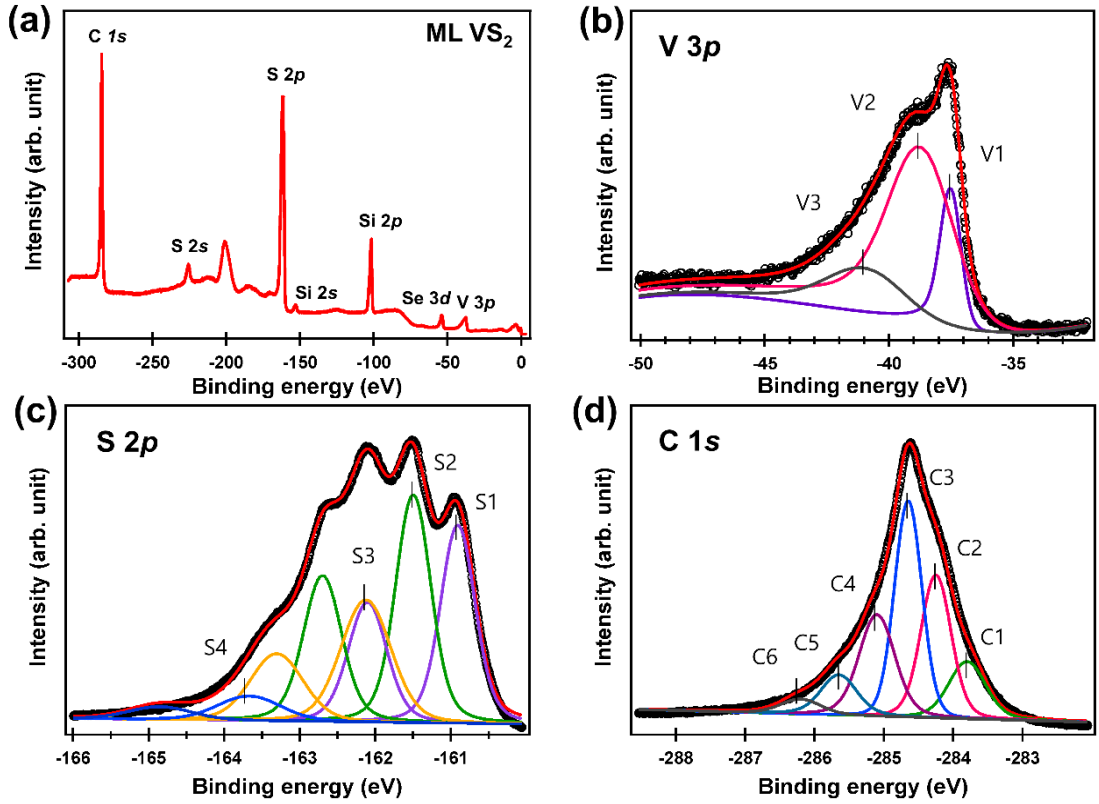

Figure 2. XPS spectra of the ML VS$_2$ at normal emission (a) wide scan, (b) V 3$p$, (c) S 2$p$, and (d) C 1$s$.

Fig. 3(a) shows the contour map of FS in the ML VS$_2$ sample at 20 K. The hexagonal BZ (yellow dashed line) corresponding to VS$_2$ lattice constant a = 3.26 Å, is smaller than the BZ of graphene (white dot-dashed lines) with Dirac cone intensity at the $K_G$ points. The FS map exhibits sixfold symmetry with six elliptic contours centered at the $M$ points. The elliptic contour (green dashed guideline) resembles the theoretically calculated FS of 1T phase ML VS$_2$, although the 1H phase has two concentric bands centered at the $\Gamma$ point [46]. The FS contour with rather rounded side sections is quite different from the FSs of ML VSe$_2$ and VTe$_2$, both of which show straightened side sections with nearly perfect FS nesting condition [33,37,47]. Although many theoretical calculations show that the ML VS$_2$ has 1H phase with spin-splitting bands with room-temperature ferromagnetic ground state, we could not detect neither the FS of 1H phase or spin-splitting of FS below ferromagnetic Curie temperature [46,48].

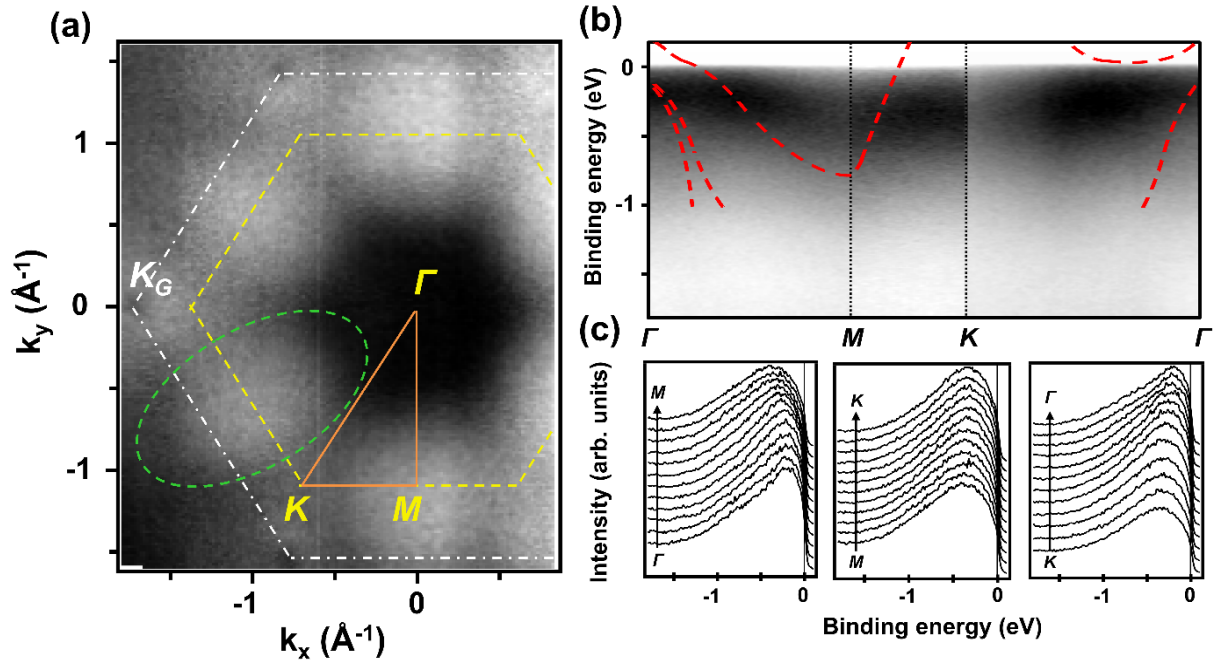

Figure 3. ARPES of ML VS$_2$. (a) Fermi surface map with hexagonal BZ of VS$_2$ (yellow dashed line) and that of BLG (white dot-dashed lines). (b) Intensity map along the $\Gamma$-$M$-$K$-$\Gamma$ (orange lines in (a)) compared with the DFT band structure from the reference [49]. (c) Energy distribution curves along $\Gamma$-$M$, $M$-$K$, and $K$-$\Gamma$ extracted from (b).

Figs. 3(b,c) shows the intensity map near $E_F$ and corresponding energy distribution curves (EDC) measured along the $\Gamma$-$M$-$K$-$\Gamma$ directions (orange lines in Fig. 3(a)), respectively. Most intensities are located at near $E_F$, attributed to V 3$d$ orbital contributions[7,48], which are displayed clearly in the EDC profiles (Fig. 3(c)). The EDC is sharp with long energy tail at the $\Gamma$ point. Along both $\Gamma$-$M$ and $\Gamma$-$K$ directions, the EDC becomes broadened with peak shift. The valence band dispersion located close to $E_F$ is similar to its bulk counterpart[7]. Although several theoretical calculations show that the ML VS$_2$ with 1T phase possesses spin-polarized band structures, we could not find well-comparing band structure within our measurements. Instead, we choose a density-functional theory calculation with undistorted ML 1T-VS$_2$ as a comparison with our band dispersion data, shown as red dashed lines in Fig. 3(b) [49]. The calculated band structure is quite well overlapped with the ARPES data along the $\Gamma$-$M$, but there is certain deviation along the $K$-$\Gamma$. Note that the calculated dispersion is about the undistorted 1T structure and additional CDW distortion introduced significant intensities just below $E_F$ along the $K$-$\Gamma$. This additional feature due to the CDW distortion further support the experimental intensity map, so that we that the ML VS$_2$ shows the band structure of 1T phase with some degrees of structure distortion associated with CDW.

Figure 4(a) shows energy dispersion curves near $E_F$ at the $\Gamma$ point for varied temperature, demonstrating evolution of CDW gap. Positions of the Fermi cutoff is gradually shifted toward lower binding energy when cooling down. In order to analyze the gap evolution, we fitted the energy gap size as a function of temperature following the Bardeen−Cooper−Schrieffer (BCS) formula, $\Delta^2(T) - \Delta^2(T_C) \propto (T_C - T)$, where $\Delta$ is the gap size and $T_C$ is CDW transition temperature, as plotted in Fig. 4(b) [50,51]. Following the linear fitting (dashed line), $T_C$ is estimated to be ~400 K. In single crystal bulk 1T-VS$_2$, CDW occur below 305 K with additional new periodicity demonstrated by electron diffraction pattern, temperature-dependent resistivity and magnetic susceptibility [7]. Such enhancement of $T_C$ in ML VS$_2$ is consistent with ML VSe$_2$, VTe$_2$, and TiTe$_2$, where 2D confinement, interfacial strain, or charge transfer are considered as major driving mechanism for stabilizing CDW in ultrathin geometries [33,37,52–55]. Considering different origins of CDW, it is of interest to investigate the influences of film thickness or different substrate choices on the electronic structures as for future studies [23,56]. In addition, FS of the bulk VS$_2$ has threefold symmetry different from the sixfold symmetric

FS of ML VS$_2$, as shown in Fig. 3(a). Although similar threefold symmetry features in FS are observed in bulk 1T-VSe$_2$[57], the ML VSe$_2$ show sixfold symmetric FS without any further band folding[33]. Such differences are understood as mainly 2D nature of ML CDW compounds, different from simple projection of bulk CDW, in addition to interactions with substrates [19,33,58].

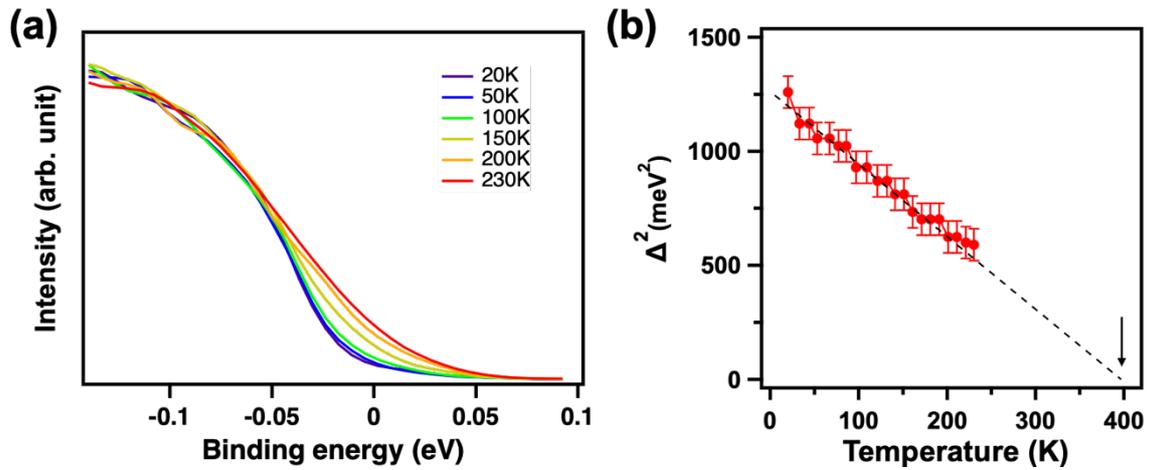

Figure 4. Temperature dependence of CDW gap of ML VS$_2$. (a) Energy dispersive curves at the $\Gamma$ point near Fermi energy, (b) Squared energy gap as a function of temperature along with linear fitting (black dashed line).

Still, there are continuation of arguments about the magnetic interactions in ML VX$_2$ (X = S, Se, Te) from both theoretical and experimental aspects. It's worth to note our observation in conjunction with the recent reports in VS$_2$ and VSe$_2$. Several x-ray magnetic circular dichroism (XMCD) studies reported that the ML VSe$_2$ possess either zero or frustrated spin ordering [24,59]. We note that recent spin-resolved ARPES measurements show that bulk VSe$_2$ has a negligible signal of spin polarization at near Fermi energy but a sizable spin polarization of Se 4$p$ state [60]. It is of great importance to unveil spin polarization in the ML VS$_2$ as well as ML VSe$_2$. Since the intrinsic ferromagnetism is recently revealed in V$_5$Se$_8$ phase, one of polymorph phase of $VSe_x$, stabilized in few layer thicknesses from MBE synthesis, it is also of great interest to further investigate intrinsic magnetic properties in other polymorph phase of $VS_x$, such as $V_5S_8$, possibly using XMCD [61,62]. In addition, CDW materials accompany phase transitions with electrical and structural reconstructions, so that CDW materials have large potential for CDW-based nonvolatile memory, oscillator, photosensor, and ultrafast switch [10,11,63]. Especially vanadium chalcogenides demonstrate variation of CDW transition temperatures depending on film thickness and stoichiometry, it is of interest to apply them for CDW-based devices near room temperature. Finally, vanadium chalcogenides attract much attention as metallic TMDCs for their surface reactivity with gases and ions for next-generation renewable energy applications [36,64]. One important approach is to analyze the valence band structure change along with the various chemical states under the ambient conditions for understanding the efficient catalytic electrode performances. Further systematic studies of core level states in the epitaxial films would be desired under ambient-pressure environments, such as ambient-pressure XPS [65,66].

## IV. CONCLUSION

In summary, we performed ARPES study of electronic structure in an epitaxial ML $VS_2$ grown on bilayer graphene by MBE. ML $VS_2$ is aligned to the graphene with the in-plane lattice parameter similar to its bulk value. We find that ML $VS_2$ has Fermi surface with six elliptic bands centered at the M point, confirming the theoretical calculation of 1T phase. Interestingly, the elliptic bands present rather round shape, different from the rather straight sides observed in both ML $VSe_2$ and ML $VTe_2$. Furthermore, it is confirmed that the temperature-dependence of CDW gap predicts the CDW transition temperature above room temperature. Our findings provide the band structure information in single-layer $VS_2$ as an important building block for establishing next-generation electronics and renewable energy applications.

## ACKNOWLEDGEMENTS

This work was supported by the National Research Foundation (NRF) grants funded by the Korean government (No. NRF-2019K1A3A7A09033389 and NRF-2020R1A2C200373211). The Advanced Light Source is supported by the Director, Office of Science, Office of Basic Energy Sciences, of the U.S. Department of Energy under Contract No. DE-AC02-05CH11231.